\documentclass[preprint,showpacs,preprintnumbers,amsmath,amssymb]{revtex4}

\usepackage{graphicx}
\usepackage{dcolumn}
\usepackage{bm}

\usepackage{feynmf}
\usepackage{slashed}

\usepackage[utf8]{inputenc}

\begin{document}

\title{ Dark Matter Relic Abundance and Light Sterile Neutrinos }

\author{Yi-Lei Tang}
\thanks{tangyilei15@pku.edu.cn}
\affiliation{Center for High Energy Physics, Peking University, Beijing 100871, China}


\author{Shou-hua Zhu}
\thanks{shzhu@pku.edu.cn}
\affiliation{Institute of Theoretical Physics $\&$ State Key Laboratory of Nuclear Physics and Technology, Peking University, Beijing 100871, China}
\affiliation{Collaborative Innovation Center of Quantum Matter, Beijing 100871, China}
\affiliation{Center for High Energy Physics, Peking University, Beijing 100871, China}

\date{\today}

\begin{abstract}

In this paper, we calculate the relic abundance of the dark matter particles when they can annihilate into sterile neutrinos with the mass $\lesssim 100 \text{ GeV}$ in a simple model. Unlike the usual standard calculations, the sterile neutrino may fall out of the thermal equilibrium with the thermal bath before the dark matter freezes out. In such a case, if the Yukawa coupling $y_N$ between the Higgs and the sterile neutrino is small, this process gives rise to a larger $\Omega_{\text{DM}} h^2$ so we need a larger coupling between the dark matter and the sterile neutrino for a correct relic abundance.

\end{abstract}
\pacs{}

\keywords{dark matter, relic abundance, sterile neutrino}

\maketitle
\section{Introduction}

The weakly interacting massive particles (WIMPs) are considered as the candidates of the dark matter (For a review, see Ref.~\cite{ParticleDarkMatter}). In this scenario, the dark matter particles are produced in the thermal bath of the early universe, then freeze out from the plasma as the temperature drops. It is well-known that the observed dark matter's relic abundance requires its  thermally averaged annihilation cross section $\langle \sigma v \rangle = 2\text{-}3 \times 10^{-26} \text{ cm}^3/\text{s}$ at the freezing-out temperature $T \sim \frac{m_{\chi}}{20}$, which is roughly the typical cross section of the weak interaction. This coincidence is called the "WIMP miracle".

Calculations of the relic abundance of the dark matter involve the Boltzmann equation (For 	 derivation, see Ref.~\cite{TheEarlyUniverse}). In the case of the WIMP dark matter, some hypotheses are adopted in order to simplify the equation. One important hypothesis is that the annihilation products of the dark matter fall in thermal equilibrium with the thermal bath rapidly. This is true when the dark matter mainly annihilates into the standard model (SM) particles. However, in many new physics models, the dark matter might mainly annihilate into other beyond-SM particles. In this case, whether this hypothesis is valid needs to be carefully examined.

In the Type I see-saw model \cite{SeeSaw1, SeeSaw2, SeeSaw3, SeeSaw4, SeeSaw5}, the right-handed neutrinos ($N$) couple with the left-handed neutrinos $l^{\pm, 0}$ through the Higgs fields $H$. After the Higgs field acquires a vacuum expectation value (VEV), the majorana mass terms of the left-handed neutrino arise through the Type I See-saw Mechanisms. In the early universe, if there is no extra sector, the main processes that can generate the right-handed neutrinos are the decay and the inverse decay of the right-handed neutrinos and the Higgs bosons (For an example of calculations, see \cite{Leptogenesis_NonThermal}). In the simplest Type I See-saw Mechanisms, if the mass of the right-handed neutrino is approximately 100 GeV, the Yukawa coupling constants of the $N$-$l^{\pm, 0}$-$H$ couplings $y_{N}$ should be smaller than $\sim 10^{-6}$ for the correct left-handed neutrino masses. However, it is too small for the right-handed neutrinos to reach in thermal equilibrium with the thermal bath. Although there are some models\cite{Inverse1, Inverse2, Inverse3, Inverse4, PseudoDirac1, PseudoDirac2, PseudoDirac3}, e.g., the inverse see-saw model, or the linear see-saw model, that can result in a larger $y_{N} \sim 0.01$,  as the temperature drops, the thermal-averaged production rates of the sterile neutrinos $\Gamma_{P} \propto e^{-\frac{m_N}{T}}$ drops rapidly and then the sterile neutrinos decay out of equilibrium.

Ref.~\cite{Secluded1, Secluded2, Secluded3} calculate a general case of the secluded dark matter, in which the annihilation products fall out of thermal equilibrium from the thermal bath. Specifically, in the literature, there are some models that the dark matter can annihilate into light sterile neutrinos \cite{BML1, BML2, BML3, BML4, NPortal1, NPortal2, NPortal3, NPortal4, FromCao}, unlike the models that the sterile neutrino itself plays the role of the dark matter \cite{AccRequired1, AccRequired2, RequiredCite}. However, as we have mentioned before, the sterile neutrinos might not be in thermal equilibrium with the SM particles in the early universe. Thus, the traditional calculations of the relic abundance might be unreliable and the standard Boltzman equation(s) should be modified. In this paper, in order to calculate these non-thermal effects, we rely on a simple model based on Ref.~\cite{NPortal2}. We focus on the case that the masses of the dark matter and the right-handed neutrinos are less than $\sim 100 \text{ GeV}$ and the dark matter particles only annihilate to right-handed neutrinos. In this case, the sterile neutrino mainly decays through the three-body final state channels. As we have calculated in Ref.~\cite{MyPaper}, this scenario can perfectly explain the gamma-ray excess from near the galactic center. We will also show that the nonthermal effects of the right-handed neutrinos can significantly modify the relic abundance when the Yukawa coupling constant $y_{N l H}$ becomes quite small.

\section{Model Descriptions}

The model discussed in this paper contains a majorana fermion $\chi$ and a real-scalar boson $\phi$. Both these two fields are SM-singlets and are odd under a dark $Z_2$ discrete symmetry. The sterile neutrino together with the SM-fields are all even under this $Z_2$ symmetry. In this paper, we discuss two cases. In one case there is one majorana right-handed neutrino $N$, and in the other case there are a pair of pseudo dirac sterile neutrino Weyl-fields $N_{1,2}$.

In the majorana right-handed neutrino case, the general Lagrangian is given by
\begin{eqnarray}
\mathcal{L} &=& \frac{1}{2} \overline{\chi} ( i \gamma^{\mu} \partial_{\mu} - m_{\chi} ) \chi +  \frac{1}{2} \overline{N} ( i \gamma^{\mu} \partial_{\mu} - m_{N} ) N + \frac{1}{2} (\partial^{\mu} \phi \partial_{\mu} \phi - m_{\phi}^2 \phi^2)  \nonumber \\
&+& y_{\chi} \overline{\chi} N \phi + i y_{\chi 5} \overline{\chi} \gamma^5 N \phi + \frac{\lambda_{\phi}}{4!} \phi^4 + \lambda_{\phi H} \phi^2 H^{\dagger} H + (y_{N i} \overline{N} P_L l_i \cdot H + \text{h.c.}) \nonumber \\
&+& \mathcal{L}_{\text{SM}}, \label{LMajorana}
\end{eqnarray}
where $N^C = N$, $\chi^C = \chi$ are written in the Dirac four-spinor form, $l_i$, $i=1,2,3$ are the left-handed lepton doublets of the three generation, $m_{\chi, \phi, N}$ are the mass terms of the $\chi$, $\phi$, $N$; $y_{\chi, \chi 5, N i}$, $\lambda_{\phi, \phi H}$ are the coupling constants, and $l_i = \left[ \begin{array}{c} \nu_i \\ e_{L_i}^{-} \end{array} \right]$, $H = \left[ \begin{array}{c} G^+ \\ v+\frac{h + i G^0}{\sqrt{2}} \end{array} \right]$ are the left-handed lepton doublet and the Higgs doublet respectively. $G^{+}$, $G^0$ are the goldstone bosons which are eaten by the gauge bosons. $v = 174 \text{ GeV}$, and $h$ is the standard model (SM) Higgs boson with a mass of $125 \text{ GeV}$. $A \cdot B$ indicates the contraction of two $SU(2)_L$ doublets, i.e., $A \cdot B = A_i (i \sigma^2_{i j}) B_j$, where $\sigma^2$ is the second Pauli-matrix.

In Eqn.~(\ref{LMajorana}), all the $y_{\chi, \chi 5, N i}$, $\lambda_{\phi, \phi H}$ and $m_{\chi, \phi, N}$ are real numbers. In fact, $y_{\chi}$ and $y_{\chi 5}$ are respectively the real part and the imaginary part of a single coupling constant $(y_{\chi} + i y_{\chi 5})\chi_w \cdot N_w \phi + \text{h.c.}$, where $\chi_w$ and $N_w$ are the Weyl components of the $\chi$ and $N$ fields. We should note that all the complex phases in the $m_{\chi, N}$ and $y_{N i}$ can be rotated away by redefining the fields and the $y_{\chi} + i y_{\chi 5}$. However, for simplicity, in the numerical calculations we just omit the $y_{\chi 5}$ and set it to be zero.

In the pseudo-Dirac sterile neutrino case, the general Lagrangian is given by
\begin{eqnarray}
\mathcal{L} &=& \frac{1}{2} \overline{\chi} ( i \gamma^{\mu} \partial_{\mu} - m_{\chi} ) \chi +  \overline{N_D} ( i \gamma^{\mu} \partial_{\mu} - m_{N_D} ) N_D + \frac{1}{2} (\partial^{\mu} \phi \partial_{\mu} \phi - m_{\phi}^2 \phi^2) \nonumber \\
&+& (\mu_1 \overline{N_D^C} P_L N_D + \mu_2 \overline{N_D^C} P_R N_D + \text{h.c.}) + \frac{\lambda_{\phi}}{4!} \phi^4 + \lambda_{\phi H} \phi^2 H^{\dagger} H \nonumber \\
&+& (y_{\chi D} \overline{\chi} N_D \phi + i y_{\chi D 5} \overline{\chi} \gamma^5 N_D \phi + y_{N i} \overline{N} P_L l_i \cdot H + y_{N C i} \overline{N^C} P_L l_i \cdot H \nonumber \\
&+& \text{h.c.} ) + \mathcal{L}_{\text{SM}},
\end{eqnarray}
where $N_D = \left[ \begin{array}{c} N_1 \\ i \sigma^2 N_2^{*} \end{array} \right]$ is a Dirac four-spinor composed of two Majorana fields $N_1$ and $N_2$, $m_D$ is the Dirac mass term of the sterile neutrinos, and $\mu_{1, 2}$ are the Majorana mass terms, $y_{\chi D}$, $y_{\chi D 5}$, $y_{N i}$, $y_{N C i}$ are the Yukawa coupling constants which is different from the Eqn. (\ref{LMajorana}). This time, we can redefine all the fields in order for the $m_{\chi}$, $m_{N_D}$ and $y_{N i}$ to be real numbers, while the $y_{\chi D}$, $y_{\chi D 5}$, $y_{N C i}$, $\mu_1$, $\mu_2$ might be complex numbers and in general their phases cannot be rotated away. In this paper, for simplicity, we only discuss the case that all these numbers are real and $y_{\chi D 5} = 0$.

In general, $\mu_{1,2}$ and $y_{N C i}$ terms violate the lepton number and cause the mass splitting of the two components of the $N_D$. If $y_{N C i}=0$ and $\mu_{1, 2} \ll m_{N_D}$, this is called the ``inverse see-saw'', and if $\mu_{1, 2}=0$ and $y_{N C i} \neq 0$, this is called the ``linear see-saw''. Thus, the masses of the left-handed neutrinos are mainly decided by the strength of the lepton number violating terms and $y_{N i}$ can be much larger than the simplest see-saw models. Usually the smallness of the left-handed neutrino masses requires the smallness of the lepton number violating terms $\mu_{1,2}$ and $y_{N C i}$. Therefore, in the early universe, the effects of the $\mu_{1,2}$ and $y_{N C i}$ terms are negligible. Therefore, these parameters are set to zero in our study and the $N_D$ is regarded as a Dirac fermion during the calculation processes.

\section{Calculations of the Dark Matter's Relic Abundance}

The calculations of the relic abundance of the dark matter are based on the Boltzmann equations (We derive the equations in this paper according to Ref.~\cite{LeptogenesisInt, Leptogenesis_NonThermal}). In principle, a full solution to the Boltzmann equations should involve the evolutions to the distribution functions of the particles. However, we assume that the elastic scatterings are fast enough as the particles in the thermal bath can maintain kenitic equilibrium as usual. For simplicity, we only consider the case that $m_{\phi} > m_{\chi} + m_{N_{(D)}}$. In the Majorana right-handed neutrino case, the Boltzmann equations are given by
\begin{eqnarray}
s H z \frac{d Y_{\chi}}{d z} &=& - \langle \sigma v \rangle_{\chi \chi \rightarrow N N} Y_{\chi eq}^2 s^2 \left( \frac{Y_{\chi}^2}{Y_{\chi eq}^2} - \frac{Y_N^2}{Y_{N eq}^2} \right) - \langle \sigma v \rangle_{\chi \chi \rightarrow \phi \phi} Y_{\chi eq}^2 s^2 \left( \frac{Y_{\chi}^2}{Y_{\chi eq}^2} - \frac{Y_{\phi}^2}{Y_{\phi eq}^2} \right) \nonumber \\
&-& \langle \sigma v \rangle_{\chi \phi \rightarrow \text{allSM}} s^2 (Y_{\chi} Y_{\phi} - Y_{\chi eq} Y_{\phi eq}) - \bar{\Gamma}_{\phi \rightarrow \chi N} Y_{\phi eq} s \left( \frac{Y_{\chi} Y_{N}}{Y_{\chi eq} Y_{N eq}} - \frac{Y_{\phi}}{Y_{\phi eq}} \right), \nonumber \\
s H z \frac{d Y_{\phi}}{d z} &=& -\langle \sigma v \rangle_{\phi \phi \rightarrow N N} Y_{\phi eq}^2 s^2 \left( \frac{Y_{\phi}^2}{Y_{\phi eq}^2} - \frac{Y_N^2}{Y_{N eq}^2} \right) - \langle \sigma v \rangle_{\phi \phi \rightarrow \chi \chi} Y_{\phi eq}^2 s^2 \left( \frac{Y_{\phi}^2}{Y_{\phi eq}^2} - \frac{Y_{\chi}^2}{Y_{\chi eq}^2} \right) \nonumber \\
&-& \langle \sigma v \rangle_{\phi \phi \rightarrow \text{allSM}} s^2 ( Y_{\phi}^2 - Y_{\phi eq}^2 ) - \langle \sigma v \rangle_{\chi \phi \rightarrow \text{allSM}} s^2 (Y_{\chi} Y_{\phi} - Y_{\chi eq} Y_{\phi eq}) \nonumber \\
&-& \bar{\Gamma}_{\phi \rightarrow \chi N} Y_{\phi eq} s \left( \frac{Y_{\phi}}{Y_{\phi eq}} - \frac{Y_{\chi} Y_{N}}{Y_{\chi eq} Y_{N eq}} \right). \nonumber \\
s H z \frac{d Y_N}{d z} &=& -\langle \sigma v \rangle_{N N \rightarrow \chi \chi} Y_{N eq}^2 s^2 \left( \frac{Y_N^2}{Y_{N eq}^2} - \frac{Y_{\chi}^2}{Y_{\chi eq}^2} \right) - 2 \langle \sigma v \rangle_{N N \rightarrow \phi \phi} Y_{N eq}^2 s^2 \left( \frac{Y_N^2}{Y_{N eq}^2} - \frac{Y_{\phi}^2}{Y_{\phi eq}^2} \right) \nonumber \\
&-& \bar{\Gamma}_{N} s (Y_N - Y_{N eq}), \label{Boltzmann}
\end{eqnarray}
where the $Y_A = \frac{n_A}{s}$ is the actual number of the constituent $A$ per-comoving-volume, and the $Y_{A eq} = \frac{n_{A eq}}{s}$ is the equilibrium number of the constituent $A$ per-comoving-volume, $n_{A (eq)}$ is the (equilibrium) number density of the constituent $A$, s is the entropy density, $z=\frac{m_\chi}{T}$, and $T$ is the temperature, $H$ is the Hubble constant. $\langle \sigma v \rangle_{AB \rightarrow CD}$ is the thermally averaged cross section times velocity
\begin{eqnarray}
\langle \sigma v \rangle_{A B \rightarrow C D} = \frac{1}{(1+\delta_{CD}) n_A n_B} \frac{g_A g_B T}{32 \pi^4} \int ds^{\prime} s^{\prime \frac{3}{2}} K_1 \left( \frac{\sqrt{s^{\prime}}}{T} \right) \lambda \left( 1, \frac{m_A^2}{s^{\prime}}, \frac{m_B^2}{s^{\prime}} \right) \sigma_{A B \rightarrow C D} (s^{\prime}),
\end{eqnarray}
where $\delta_{CD} = 1(0)$ if $C$ and $D$ are identical(different) particles, $g_A$ and $g_B$ are the degrees of freedoms of particle $A$ and $B$, $K_1$ is a Bessel function, $\sigma_{A B \rightarrow C D}(s^{\prime})$ is the cross section of the process $A B \rightarrow C D$ with the total energy in the center of mass frame is $\sqrt{s^{\prime}}$.

The definition of the $\bar{\Gamma}_{\phi \rightarrow \chi N}$ is given by
\begin{eqnarray}
\bar{\Gamma}_{\phi \rightarrow \chi N} = \frac{K_1(\frac{m_{\phi}^t}{T})}{K_2(\frac{m_{\phi}^t}{T})} \Gamma_{\phi \rightarrow \chi N},
\end{eqnarray}
where $m_{\phi}^t$ is the thermal mass of $\phi$ which will be defined later. The $\bar{\Gamma}_{N}$ is a little bit complicated. We need to consider the decay/inverse-decay processes $N \rightarrow h^{\pm 0} l$ or $h^{\pm 0} \rightarrow l N$. However, as the temperature drops below the electroweak symmetry breaking (EWSB) critical temperature $T_c$, we should considers the processes $N \leftrightarrow W^{\pm}/Z/h$. In this paper, we adopt the approximation method described in Ref.~\cite{Liumangfa} to calculate the $N \rightarrow h^{\pm 0} l$ or $h^{\pm 0} \rightarrow l N$ with all the four states of the Higgs doublets having the Higgs boson mass $m_h(T)$ below $T_c$. If $m_N > m_h(T)$,
\begin{eqnarray}
\tilde{\Gamma}_N = \frac{K_1 (\frac{m_N}{T})}{K_2 (\frac{m_N}{T})} \Gamma_{N \rightarrow H + l},
\end{eqnarray}
while $m_N < m_h(T)$,
\begin{eqnarray}
\tilde{\Gamma}_N = \frac{Y_{H eq}}{Y_{N eq}} \frac{K_1 (\frac{m_h(T)}{T})}{K_2 (\frac{m_h(T)}{T})} \Gamma_{H \rightarrow N l}. \label{InverseNDecay}
\end{eqnarray}
However, when $T \ll m_h(0\text{ GeV}) = 125 \text{ GeV}$, $\tilde{\Gamma}_N$ is severely suppressed by a factor of $e^{\frac{-2 m_h + m_N}{T}}$. Once it is less then the $\frac{K_1 (\frac{m_N}{T})}{K_2 (\frac{m_N}{T})} \Gamma_{N \rightarrow h^*/W^*/Z^* l}$ where $\Gamma_{N \rightarrow h^*/W^*/Z^* l}$ is calculated at the zero temperature, we set
\begin{eqnarray}
\tilde{\Gamma}_N = \frac{K_1 (\frac{m_N}{T})}{K_2 (\frac{m_N}{T})} \Gamma_{N \rightarrow h^*/W^*/Z^* l}
\end{eqnarray}
in order to let the right-handed neutrino decay.

We calculate the thermal masses in the following procedures. The thermal effects on the fermions are neglected. As for the scalar bosons, the effective potential in a finite temperature $T$ is \cite{FiniteT1, FiniteT2, FiniteTLeptogenesis1, FiniteTLeptogenesis2}
\begin{eqnarray}
V_{\text{eff}} (h, \phi, T) = \frac{\lambda}{4} h^4 + \frac{1}{2} (\mu^2 + c T^2) h^2 + \frac{\lambda_{\phi}}{4!} \phi^4 + \frac{\lambda_{\phi h}}{2} \phi^2 h^2 + \frac{1}{2}(m_{\phi}^2 + c_{\phi} T^2) \phi^2,
\end{eqnarray}
where $\lambda$ is the self-interacting coupling constant of the Higgs boson, $\mu^2<0$ is the mass term at zero temperature of the Higgs potential. The definition of the $c$ and $c_{\phi}$ is given by
\begin{eqnarray}
c &=& \frac{1}{16} (g_1^2 + 3 g_2^2 + 4 y_t^2 + 4 \frac{m_h^2}{v^2} ) + \frac{\lambda_{\phi h}}{12}, \nonumber \\
c_{\phi} &=& \frac{1}{12} (2 y_{\chi}^2 + 2 y_{\chi 5}^2 + \frac{\lambda_{\phi}}{2} + 4 \lambda_{\phi h}). \label{CCPhi}
\end{eqnarray}
The critical temperature $T_c$ of the EWSB is
\begin{eqnarray}
T_c = \sqrt{\frac{-\mu^2}{c}}.
\end{eqnarray}
Then the temperature dependent masses of the Higgs boson and the scalar $\phi$ are given by
\begin{eqnarray}
m_h(T) &=& \left\lbrace
\begin{array}{cc}
\sqrt{\mu^2 + c T^2}, & ~(T>T_c) \\
\sqrt{-2(\mu^2 + c T^2)}, & ~(T<T_c)
\end{array} \right. , \nonumber \\
m_{\phi}(T) &=& \lambda_{\phi h} v(T)^2 + m_{\phi}^2 + c_{\phi} T^2,
\end{eqnarray}
where
\begin{eqnarray}
v_{T} = \sqrt{-\frac{\mu^2+c T^2}{\lambda}}.
\end{eqnarray}

As for the pseudo-Dirac sterile neutrino case, all the $N$'s in the above formulas should be replaced with $N_D$ and $\bar{N}_D$. One needs to note that the condition of $Y_{\bar{N}_D} = Y_{N_D}$ always holds and a summation over the particle and anti-particle should be considered. That is to say, in (\ref{Boltzmann}), $\langle \sigma v \rangle_{A A \rightarrow N N}$ should be replaced with $\langle \sigma v \rangle_{A A \rightarrow N_D N_D} + \langle \sigma v \rangle_{A A \rightarrow \bar{N}_D N_D} + \langle \sigma v \rangle_{A A \rightarrow \bar{N}_D \bar{N}_D}$, $\bar{\Gamma}_{\phi \rightarrow \chi N}$ should be replaced with $\bar{\Gamma}_{\phi \rightarrow \chi N_D} + \bar{\Gamma}_{\phi \rightarrow \chi \bar{N}_D}$, $\langle \sigma v \rangle_{N N \rightarrow A A}$ should be replaced with $\langle \sigma v \rangle_{N_D N_D \rightarrow A A} + \langle \sigma v \rangle_{N_D \bar{N}_D \rightarrow A A}$. In (\ref{CCPhi}), the $2 y_{\chi}^2 + 2 y_{\chi 5}^2$ should also be replaced with $4 y_{\chi D}^2 + 4 y_{\chi D 5}^2$.

\section{Numerical Solutions and Results}

To solve the differential equations (\ref{Boltzmann}), we use the ready-made function \cite{Stiff1, Stiff2} embeded in the micrOMEGAs \cite{micrOMEGAs} for computing the stiff equations Eqn.~(\ref{Boltzmann}). We also use the CalcHEP \cite{CalcHEP} embeded in the micrOMEGAs to calculate the $\langle \sigma v \rangle (s)$ and the widths of the particles. The model file is implemented and output by FeynRules \cite{FeynRules}. We adopt the $g_*$ and $g_{*S}$ implemented in the micrOMEGAs in order to calculate the Hubble constant $H = 1.66 \sqrt{g_*} \frac{T^2}{M_{pl}}$, and $s = \frac{2 \pi^2}{45} g_{*S} T^3$. Here $M_{pl} = 1.22 \times 10^{19} \text{ GeV}$ is the Planck energy.

If the $\lambda_{\phi h}$ is too small, and when $T \gg m_{\phi}$, $\phi$ becomes in thermal equilibrium with the SM particles through the $\phi \phi \leftrightarrow H H$ interactions. The $\chi$ and $N_{(D)}$ then fall into thermal equilibrium through the $\phi$-portal. However, once the temperature drops below the mass of the $\phi$, the number density of $\phi$ rapidly becomes so small that the $\chi$ and $N_{(D)}$ decouple from the thermal bath altogether. Finally, $\chi$, $N_{(D)}$ decouple from each other.

\begin{figure}
  \includegraphics[width=3.2in]{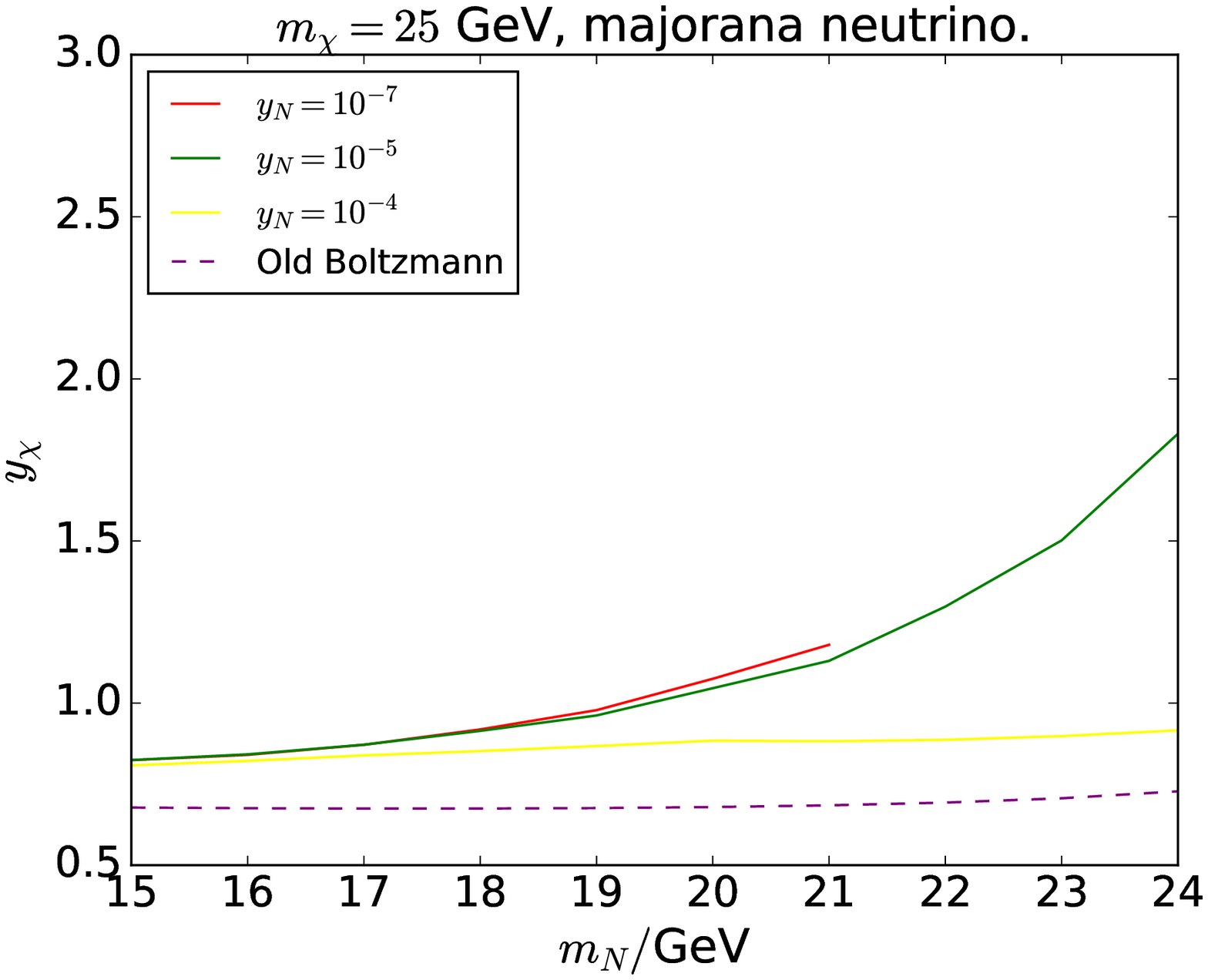}
  \includegraphics[width=3.2in]{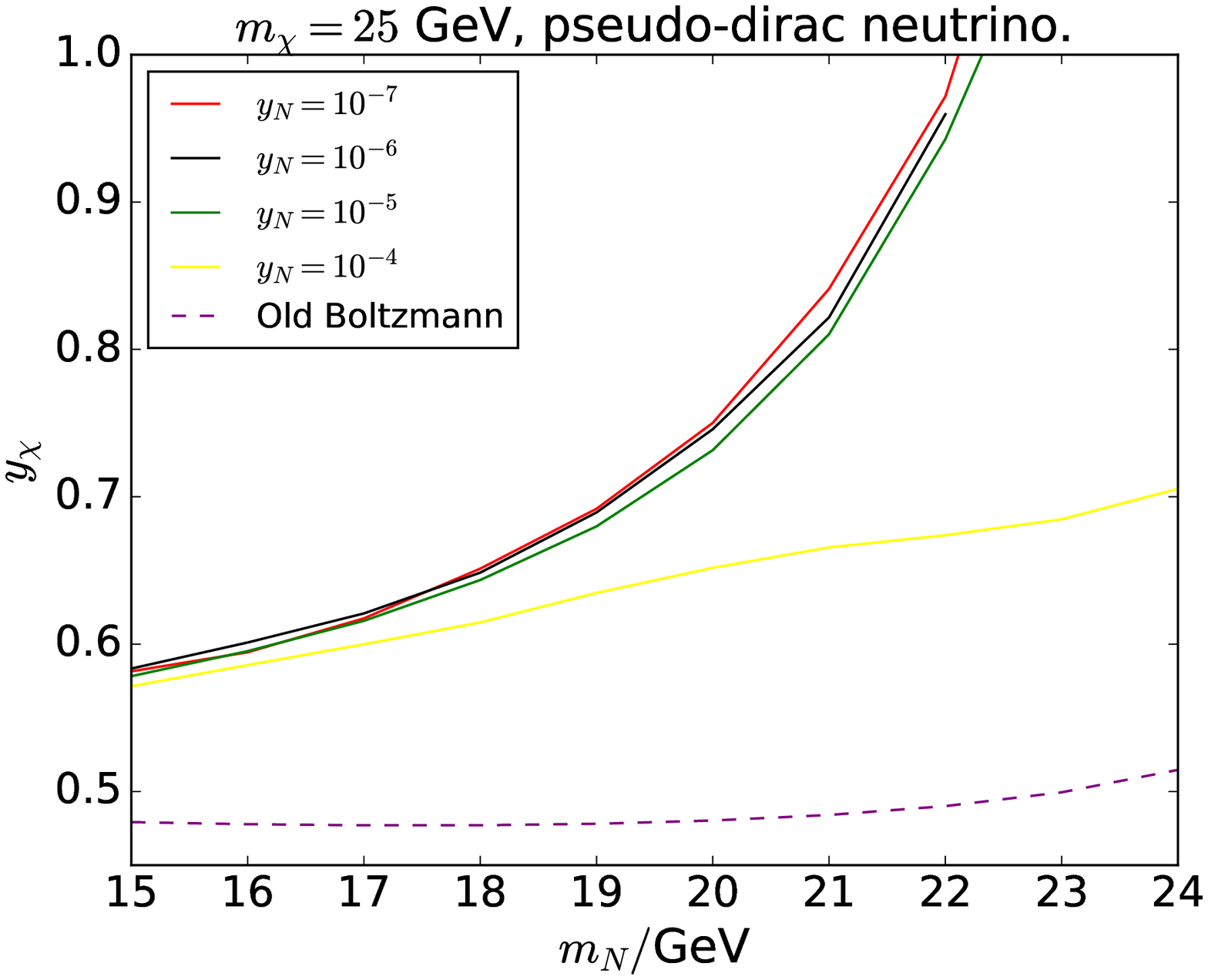}
  \caption{$m_{\chi} = 25 \text{ GeV}$ for the majorana sterile neutrino (left panel) and pseudo-dirac sterile neutrino (right panel) case. }
  \label{Chi25}
\end{figure}

\begin{figure}
  \includegraphics[width=3.2in]{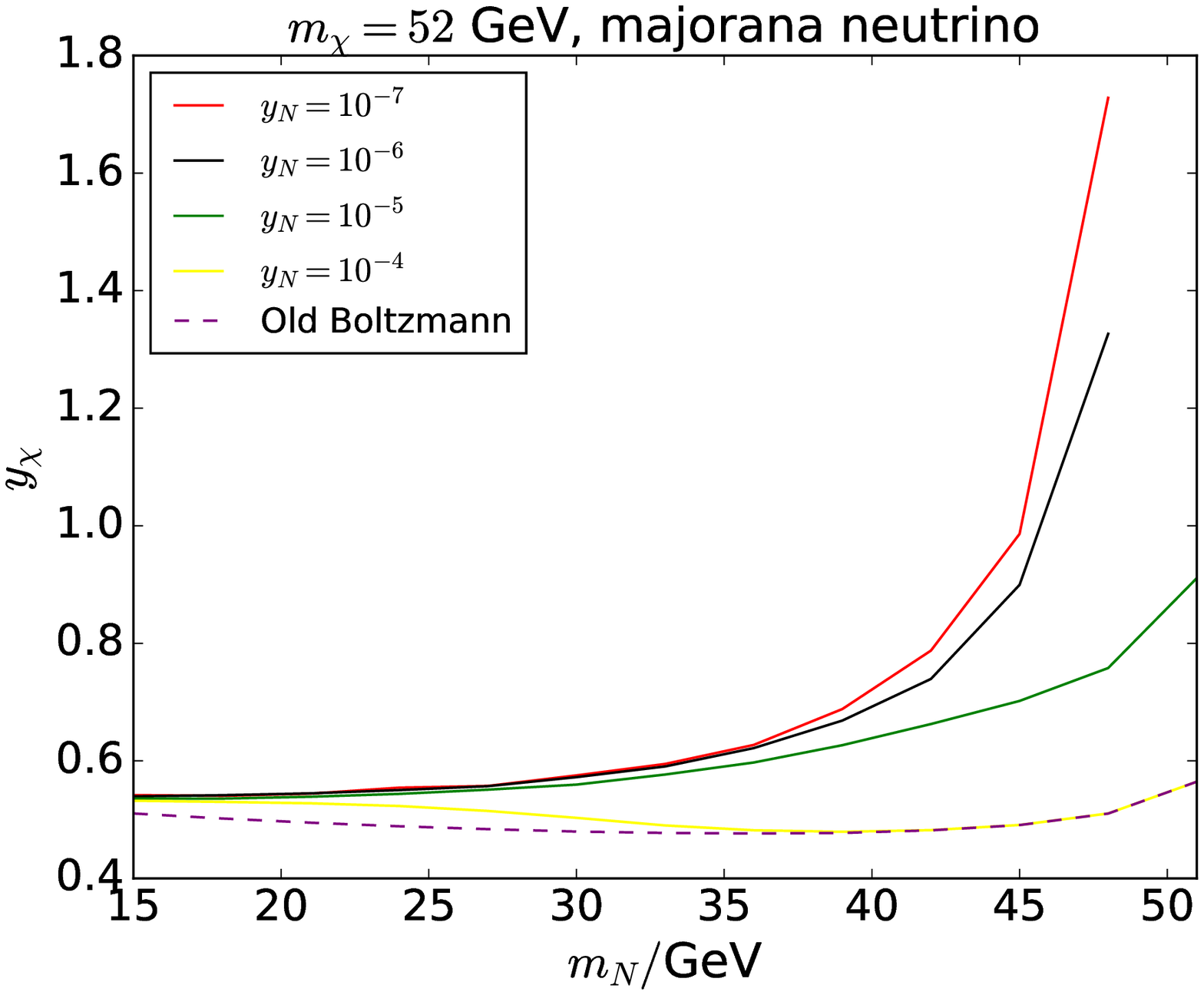}
  \includegraphics[width=3.2in]{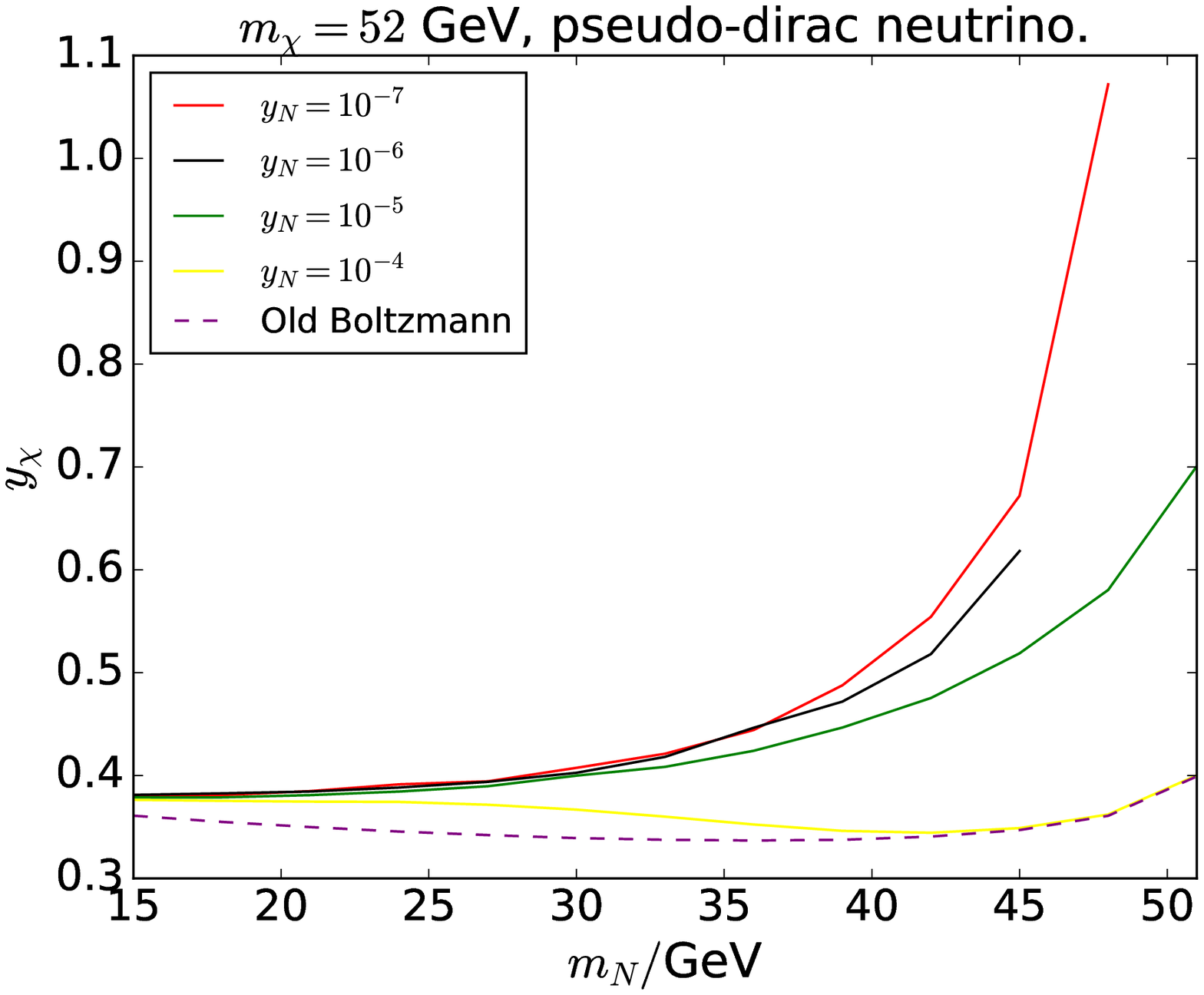}
  \caption{$m_{\chi} = 52 \text{ GeV}$ for the majorana sterile neutrino (left panel) and pseudo-dirac sterile neutrino (right panel) case.}
  \label{Chi50}
\end{figure}

\begin{figure}
  \includegraphics[width=3.2in]{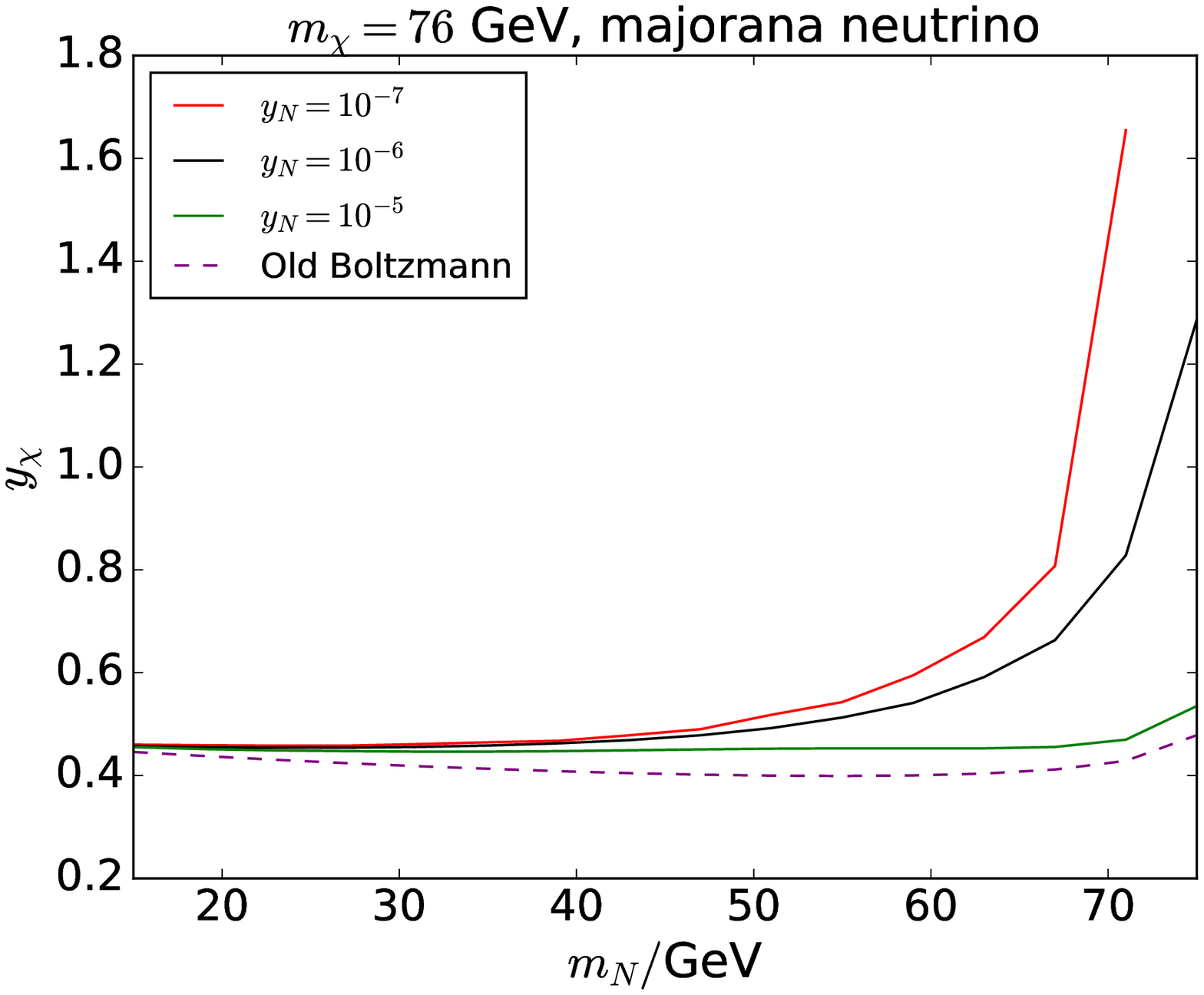}
  \includegraphics[width=3.2in]{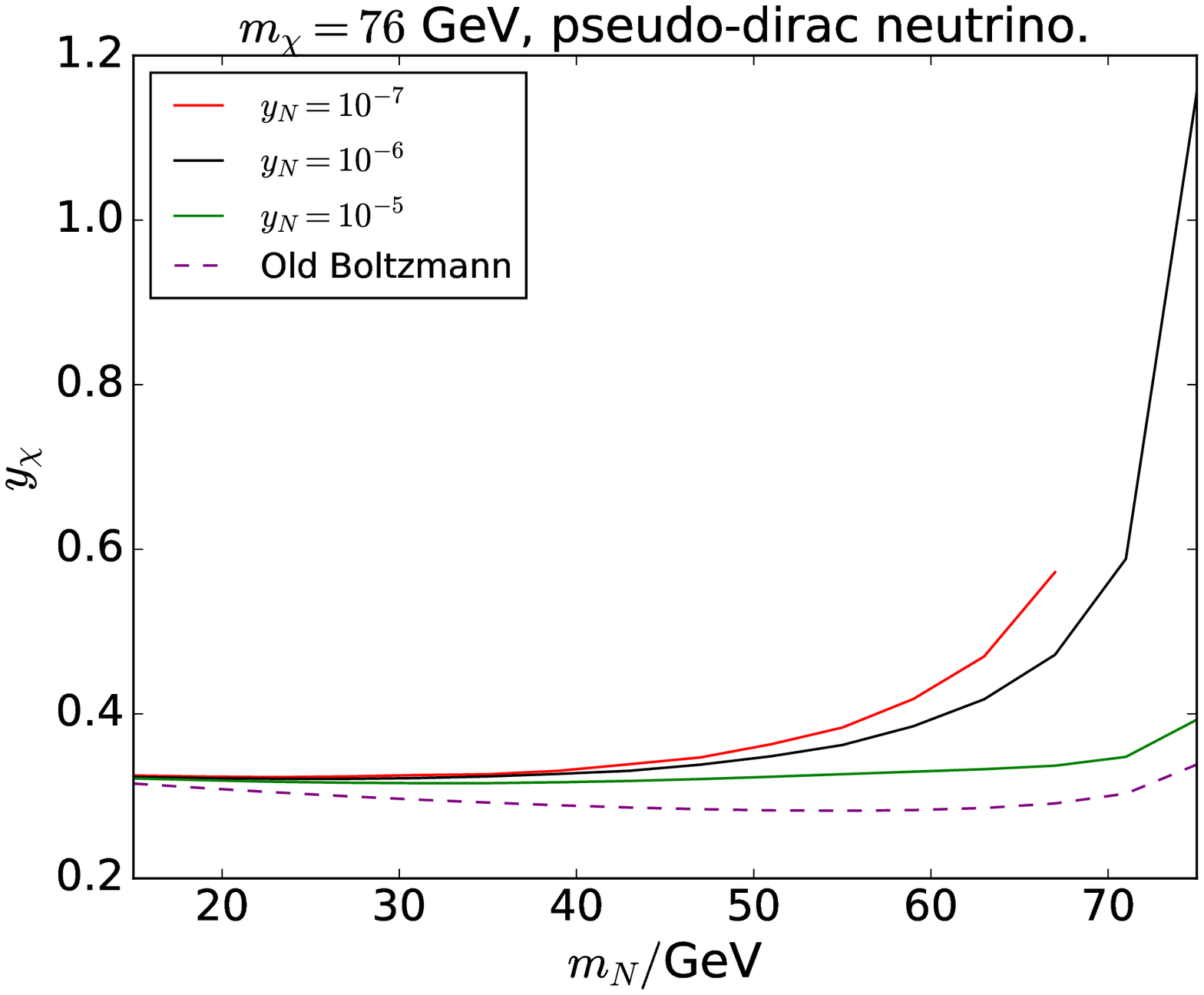}
  \caption{$m_{\chi} = 76 \text{ GeV}$ for the majorana sterile neutrino (left panel) and pseudo-dirac sterile neutrino (right panel) case.}
  \label{Chi75}
\end{figure}

During the calculations, we simplify the Eqn.~(\ref{Boltzmann}) by eliminating all the terms involving $Y_{\phi}$ once $Y_{\phi} < 0.01 Y_{\chi}$. In order to present our result, we fix $m_{\phi} = 180 \text{ GeV}$, $\lambda_{\phi} = 0.5$ and $\lambda_{\phi H}=0.45$, $y_{N2}=y_{N3}=0$. We plot our results on the $m_{N_D}$-$y_{\chi}$ plane in the different combinations of the values of $y_N=y_{N 1}=10^{(-7), (-6), (-5), (-4), (-3), (-2)}$ (most of the values are far beyond the current collider bounds, for the related discussions, see Ref.~\cite{RHNCollider1, RHNCollider2, RHNCollider3, RHNCollider4, RHNCollider5, RHNCollider6, RHNCollider7, RequiredRHNLimit1, RequiredRHNLimit2, RequiredRHNLimit3, DasRequired}), and $m_{N_(D)}=25, 52, 76 \text{ GeV}$. For each $m_{N_{(D)}}$, we find one $y_{\chi}$ that results in $0.117 < \Omega_{\text{DM}} h^2 < 0.120$ \cite{PlanckResult}. For comparison, we also present the result using the traditional Boltzmann equation
\begin{eqnarray}
s H z \frac{d Y_{\chi}}{dz} = -\langle \sigma v \rangle_{\chi \chi \rightarrow N_{(D)} N_{(D)}} s^2 (Y_{\chi}^2 - Y_{\chi eq}^2).
\end{eqnarray}
The results are shown in Fig.~\ref{Chi25}, \ref{Chi50} and \ref{Chi75}. We should note that in these figures, some lines are left over since they are nearly identical with the drawn lines. When $y_N \gtrsim 10^{-3}$, the deviation from the standard calculation by the old Boltzmann equation is quite small.
Note that when $y_N \lesssim 10^{-5}$ and $m_{N_{(D)}}$ approaches $m_{\chi}$, the numerical processes of solving the Boltzmann equation become very slow, and we do not include the complete results in this case. Therefore, some lines may disappear before the most right-handed side in Fig.~\ref{Chi25}-\ref{Chi75}.

\begin{figure}
  \includegraphics[width=4in]{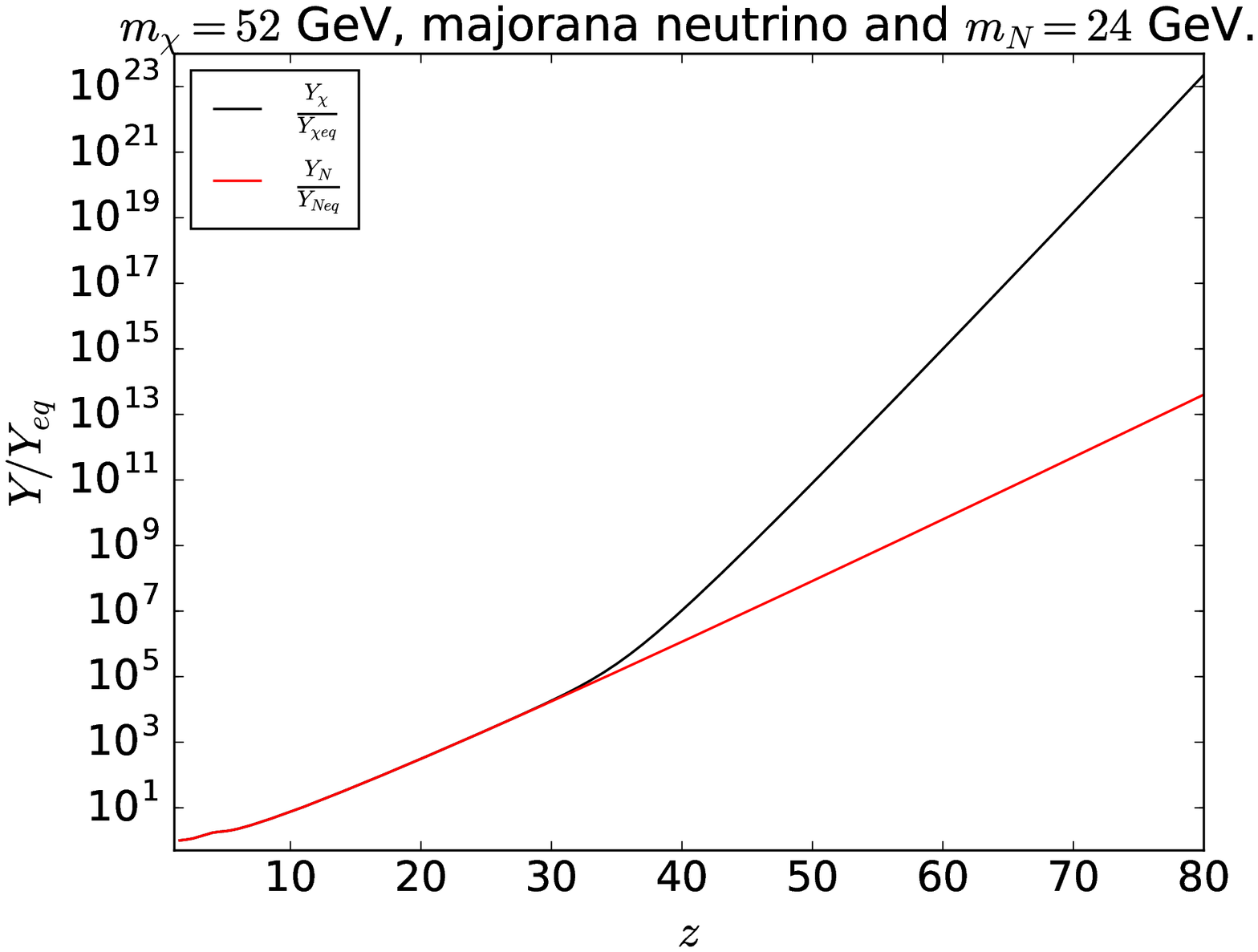}
  \caption{The $z$-evolution of $\frac{Y_{\chi}}{Y_{\chi eq}}$ and $\frac{Y_{N}}{Y_{N eq}}$ in the benchmark point that $m_N=24 \text{ GeV}$, $m_{\chi}=52 \text{ GeV}$, $y_{\chi} = 0.554$, and $y_N = 10^{-7}$.} \label{z_Evolution}
\end{figure}

To study the decoupling processes in details, we plot the $z$-evolution of $\frac{Y_{\chi}}{Y_{\chi eq}}$ and $\frac{Y_{N}}{Y_{N eq}}$ in Fig.~\ref{z_Evolution}.
We can see that before $z \lesssim 30$, the $\chi$ and the sterile neutrino together decouple from the thermal bath. As they are in thermal equilibrium with each other, $\frac{Y_{\chi}}{Y_{\chi eq}}$ traces the $\frac{Y_{N}}{Y_{N eq}}$ very well. Then  the $\chi$ and the $N$ decouple from each other and $\chi$ finally freezes out. In the region where $z>45$, although $\frac{Y_{N}}{Y_{N eq}}$ still arises as the $z$ accumulates, $y_{N}$ actually drops as $y_{N eq}$ decreases much faster. When $\chi$ and $N_{(D)}$ decouples with each other while $y_N$ is small, $Y_{N_{(D)}}$ is usually larger than $Y_{N_{(D)} eq}$, which gives rise to the $Y_{\chi}$ at the freeze-out point. That is the reason we need a larger $y_{\chi}$ to suppress the $Y_{\chi \infty} \propto \Omega_{\text{DM}} h^2$ in the case of small $y_{N} \ll 10^{-3}$ as shown in Fig.~\ref{Chi25}-\ref{Chi75}.

\section{Discussions}

In the case that the fermionic $\chi$ to be the dark matter candidate, there is no tree-level diagrams contributing to the direct detection processes. However, As has been mentioned in Ref.~\cite{NPortal1, NPortal2}, one-loop diagram will result in $\overline{\chi} (I\text{, or }i \gamma^5) \chi H^{\dagger} H$ operators, which give rise to not only the direct detection processes through exchanging a Higgs boson with the target nucleon, but also lead to the Higgs invisible decays. These effects are all suppressed by the loop factor and are proportional to $\lambda_{\phi H}^2 y_{\chi}^4$. The Ref.~\cite{NPortal2} calculated these constraints in a similar model, and the result was shown in its Fig.~2. From the left panel we can see that we do not need to worry about these contraints once $\lambda_{\phi H} \lesssim 1$. Although in this paper, we need a larger $y_{\chi}^4$ than the usual standard calculations in order to get an appropritate dark matter relic abundance, we can at least escape the constraints by assigning a smaller $\lambda_{\phi H}$ accordingly. Such an assignment usually does not affect the relic abundance of the dark matter significantly, because the main annihilation channels do not involve $\lambda_{\phi H}$.

Ref.~\cite{MyPaper} has calculated the galactic center gamma-ray excess in such kind of scenario. In Ref.~\cite{MyPaper} we have pointed out that an approximately 10-60 GeV sterile neutrino together with a heavier dark matter particle can perfectly explain the the observed spectrum. The annihilation cross section $\langle \sigma v \rangle$ is within the range $0.5$-$4 \times 10^{-26} \text{cm}^3/\text{s}$. Considering the uncertainties of the parameters of the dark matter profile, e.g., the local dark matter density $\rho_{\odot}$ which varies from 0.2-0.6 $\text{GeV}/\text{cm}^3$, $\langle \sigma v \rangle_{\text{real}} \propto \rho_{\odot}^{-2}$ can differ by one order of magnitude. In this paper, we need a larger $y_{\chi}^4$, which amplifies the $\langle \sigma v \rangle_{\text{real}} \propto y_{\chi}^4$ from the standard WIMP cross section $\sim 3 \times 10^{-26} \text{cm}^3/\text{s}$ in order for a correct dark matter relic abundance. Such an amplification factor is typically $\sim 2$-$100$ depending on the $m_{\chi, N}$ when $y_{\chi} \ll 10^{-4}$, leaving us enough room to adjust the dark matter profile parameters to fit the gamma-ray data from the galactic center.

Finally, we should note that even in the extreme case that the $y_{N}$ is as small as $10^{-7}$ which lead to a long-life right-handed neutrino, the Big-Bang neucleosynthesis (BBN) is not affected (For a review of the BBN, see the section 24 in the Ref.~\cite{PDG}, and for references therein). For example, the width $\Gamma_N \sim 10^{-17} \text{ GeV}$ when $m_N \sim 50 \text{ GeV}$, but this is still much larger compared with the Hubble constant $H \sim 10^{-22} \text{ GeV}$ at the BBN temperature $T \sim 10 \text{ MeV}$, so nearly all of the out-of-equilibrium sterile neutrino decay before they may have an impact on the BBN.

\section{Conclusions}

We have calculated the relic abundance of the dark matter particles when they annihilate into sterile neutrinos with the mass $m_N < m_{\chi} \lesssim 100 \text{ GeV}$. In the model we have relied on, the sterile neutrino will become in thermal equilibrium with the thermal bath when $T \gg m_{N_{(D)}}$, however it will decouple from the thermal bath before the dark matter freezes out if $y_N$ is small. This gives rise to a larger $\Omega_{\text{DM}} h^2$ so we need a larger coupling between dark matter and the sterile neutrino for a correct relic abundance. In the future, we will continue to dedicate ourselves in some more detailed research in such kind of scenarios.

\begin{acknowledgements}

We would like to thank Ran Ding, Weihong Zhang, Zhao-Huan Yu, Oliver Fischer, Andrew J. Long, Qinghong Cao, Xuan Chen for helpful discussions.  This work was supported in part by the Natural Science Foundation of China (Grants No.~11135003, No.~11635001 and No.~11375014), and by the China Postdoctoral Science Foundation under Grant No.~2016M600006.

\end{acknowledgements}

\newpage
\bibliography{RHNDM_Relic}
\end{document}